\documentclass{article}

\usepackage{graphicx} 
\usepackage{float}
\usepackage{authblk}

\usepackage[letterpaper,top=2cm,bottom=2cm,left=3cm,right=3cm,marginparwidth=1.75cm]{geometry}
\usepackage{algorithm}
\usepackage{algpseudocode}
\usepackage{amsmath}
\usepackage{multirow}
\usepackage{booktabs}
\usepackage{float}
\usepackage{makecell}
\usepackage{CJKutf8}
\usepackage[numbers,sort&compress]{natbib}
\usepackage{enumitem}

\usepackage[colorlinks=true, urlcolor=blue, linkcolor=red]{hyperref}

\title{Survey of HPC in US Research Institutions}

\newcommand{\equalcontrib}{\textsuperscript{*}}

\author[1]{Peng Shu\thanks{Co-first author}}
\author[1]{Junhao Chen\equalcontrib}
\author[1]{Zhengliang Liu}
\author[1]{Huaqin Zhao}
\author[1]{Xinliang Li}
\author[1]{Tianming Liu\thanks{Corresponding author}}

\affil[1]{School of Computing, The University of Georgia, Athens, GA, 30602, USA}

\begin{document}
\maketitle

\begin{abstract}
The rapid growth of AI, data-intensive science, and digital twin technologies has driven an unprecedented demand for high-performance computing (HPC) across the research ecosystem. While national laboratories and industrial hyperscalers have invested heavily in exascale and GPU-centric architectures, university-operated HPC systems remain comparatively under-resourced. This survey presents a comprehensive assessment of the HPC landscape across U.S. universities, benchmarking their capabilities against Department of Energy (DOE) leadership-class systems and industrial AI infrastructures. We examine over 50 premier research institutions, analyzing compute capacity, architectural design, governance models, and energy efficiency. Our findings reveal that university clusters, though vital for academic research, exhibit significantly lower growth trajectories (CAGR $\approx$ 18\%) than their national ($\approx$ 43\%) and industrial ($\approx$ 78\%) counterparts. The increasing skew toward GPU-dense AI workloads has widened the capability gap, highlighting the need for federated computing, idle-GPU harvesting, and cost-sharing models. We also identify emerging paradigms, such as decentralized reinforcement learning, as promising opportunities for democratizing AI training within campus environments. Ultimately, this work provides actionable insights for academic leaders, funding agencies, and technology partners to ensure more equitable and sustainable HPC access in support of national research priorities.
\end{abstract}

\section{Introduction}

\subsection{Context and Motivation}

In recent years, the scientific and technological landscape has witnessed a dramatic rise in the demand for high-performance computing (HPC), driven primarily by the exponential growth of artificial intelligence (AI) and machine learning (ML), the acceleration of data-intensive scientific research, and the emergence of complex digital twin technologies. These domains increasingly require computational capacities that exceed traditional campus infrastructures, compelling universities to reassess and scale their HPC capabilities. However, despite these rapidly evolving needs, there remains a pronounced gap between university-operated HPC resources and those available at Department of Energy (DOE) national laboratories or hyperscale AI facilities operated by technology corporations such as Google, Microsoft, and NVIDIA.

The disparity is not solely in terms of computational power; significant differences exist in system architectures, resource accessibility, and governance models. While national labs and industry players have made considerable investments in exascale computing platforms and massively parallel GPU-centric infrastructures, university HPC systems typically remain constrained by funding limitations, traditional architectural models, and fragmented governance structures. This imbalance poses a significant challenge for academic research institutions seeking to maintain competitiveness in cutting-edge research areas.

\subsection{Objectives}

The primary objective of this survey is to quantitatively and qualitatively assess the differences among university-operated HPC systems, DOE leadership-class systems, and industrial hyperscale AI infrastructures. By systematically comparing these sectors across key dimensions—such as computing scale, architectural choices, access policies, governance frameworks, and sustainability models—this work aims to highlight both existing capability gaps and potential opportunities for strategic enhancement.

Moreover, this analysis seeks to deliver actionable insights tailored explicitly to stakeholders within academia, government funding bodies, and industry collaborators. Campus leaders can utilize these insights to inform strategic planning and resource allocation decisions. Funding agencies can leverage the comparative findings to better align investments with the critical infrastructure needs of universities. Industry partners, meanwhile, may identify opportunities for meaningful collaborations and resource-sharing arrangements that benefit both academic research and industrial innovation.

Ultimately, this survey intends to foster a deeper understanding of the HPC landscape, empowering all stakeholders to make informed decisions that support the sustained growth and advancement of science and technology at the nation’s leading universities.

\section{Scope and Definitions }
In this survey, we categorize U.S. HPC resources into three sectors – University HPC, National-Lab HPC, and Industrial HPC – and define the scope of systems considered. We focus on large-scale computing installations (larger than 100 TFLOPS or larger than 100 GPUs) and outline which resources are included or excluded. Below we define each sector and the inclusion/exclusion policy, noting that the university sector is considerably less resourced compared to the national and industrial sectors.
  \subsection{Sector Categories}
    \begin{enumerate}[label*=\arabic*.]
      \item \textbf{University HPC} — campus-wide resources at the top U.S. universities
      
      University HPC refers to campus-wide or multi-institution shared computing platforms operated by major research universities (approximately the top 50 in the U.S.). These are centrally managed clusters that serve a broad range of academic researchers and students across disciplines. Such systems typically provide a few hundred to a few thousand compute nodes and may incorporate GPU accelerators, but they are modest in scale relative to national facilities. Even the most powerful university-owned supercomputers tend to peak in the range of ~0.1–10 petaflops of performance (e.g. Clemson University’s Palmetto cluster at ~0.397 PF \cite{clemson_palmetto} and Purdue’s Gautschi cluster at ~10 PF peak \cite{purdue_gautschi}). This is orders of magnitude lower than the capabilities of the other sectors. In short, campus HPC resources – while vital for university research competitiveness – remain significantly under-resourced compared to the leadership-class machines and massive industrial supercomputers described below \cite{frontier_top500}.
      
      \item \textbf{National-Lab HPC} — DOE, NSF, and other federally funded leadership systems

      National-Lab HPC encompasses the federally supported, leadership-class computing facilities that serve as national infrastructure for science and engineering. This includes the supercomputing centers at U.S. Department of Energy national laboratories (for example, Oak Ridge National Laboratory’s OLCF and Argonne National Laboratory’s ALCF) as well as NSF-funded national centers (such as those in the NSF ACCESS program) that provide HPC resources to researchers nationwide \cite{columbia_cuit_hpc}. Systems in this sector are characterized by their very high performance and scale, often ranking among the top supercomputers globally. They are designed to tackle grand-challenge simulations and large-scale data analyses, with leadership-class machines now reaching multi-petaflop to exaflop performance levels \cite{frontier_top500}. For instance, the Frontier system at Oak Ridge was the world’s first deployed exascale supercomputer, achieving 1.21 EFLOPS (\(1.21 \times 10^{18} \) FLOP) on the LINPACK benchmark \cite{frontier_top500}. These national HPC facilities typically support open science through peer-reviewed allocations and are funded by federal agencies to enable cutting-edge research across academia, government, and industry.
      
      \item \textbf{Industrial HPC} — corporate R\&D clusters and hyperscale AI infrastructures

      Industrial HPC refers to high-performance computing resources operated by corporations and hyperscale technology companies for research, development, and product infrastructure. This sector includes two main categories: (1) corporate R\&D supercomputers, such as those used in aerospace, energy, finance, or pharmaceutical companies, and (2) the massive hyperscale AI and cloud computing infrastructures run by companies like NVIDIA, Microsoft, Google, Amazon, Meta, and others. These systems are often GPU-accelerated clusters or superpods dedicated to demanding workloads (e.g. AI model training, big data analytics) and are typically proprietary or cloud-based. Notably, the scale of industrial HPC now rivals or exceeds academic centers. For example, NVIDIA’s internal Selene supercomputer achieved 63.4 petaflops on the TOP500 Linpack test (ranked 5 worldwide at the time) \cite{nvidia_selene}, making it one of the fastest commercial machines. Cloud providers have also built leadership-class systems – the Microsoft Azure “NDv5” cluster recently recorded over 560 petaflops, the highest-ranking cloud system on the TOP500 as of 2024 \cite{frontier_top500}. Moreover, hyperscale AI installations consist of thousands of GPU accelerators: Meta’s AI Research SuperCluster (RSC), for instance, deploys 760 NVIDIA DGX nodes (6,080 GPUs) for a total peak performance on the order of multiple exaflops in AI throughput \cite{azure_ndv5_top500}. In summary, the industrial sector’s HPC resources are distinguished by their vast scale and emphasis on GPU-based computing, far outstripping typical university HPC capabilities in raw power and scale of deployment.
      
    \end{enumerate}
  \subsection{Inclusions and Exclusions}
    For the present survey an HPC installation is deemed in-scope only if it satisfies at least one of two quantitative thresholds — (i) a sustained or theoretical peak performance of no less than \(10^{2}\) TFLOPS or (ii) an aggregate of one hundred or more general-purpose GPU accelerators. These limits ensure that the study captures systems of sufficient architectural scale to permit meaningful comparison across the university, national-laboratory, and industrial sectors. Conversely, we exclude short-lived or credit-based cloud allocations that lack an enduring, institutionally controlled hardware footprint; cloud resources are considered only when the underlying infrastructure is physically deployed on the premises of the host organisation (e.g., a vendor-operated supercomputer installed in a campus data centre). By surveying exclusively those facilities that meet the foregoing inclusion criteria—and omitting smaller clusters or purely virtual resources—we confine our analysis to persistent, large-scale platforms whose capabilities and governance structures are most relevant to the comparative objectives of this work.

\section{Sector Overviews }
  \subsection{University HPC Landscape (Top 50)}
    HPC is a cornerstone of research infrastructure at U.S. universities. Dozens of major research universities maintain centralized HPC centers that provide computational power for science and engineering across disciplines. In recent years, a handful of these academic supercomputers have achieved performance levels on par with the world’s fastest systems – although the very top spots are generally occupied by federal lab machines, universities continue to push the boundaries of HPC within academia. Indeed, the fastest university-based systems now reach petascale speeds, and in a few cases approach the lower rungs of the global top 50 by performance \cite{top5002023june}. At the same time, many institutions support smaller cluster environments and “condominium” arrangements (shared faculty-owned nodes) to serve the broad needs of campus researchers. This section surveys the landscape of university HPC, focusing on leading systems (approximately top-50 caliber) and general trends in the academic sector.

     A small number of university HPC centers operate machines that rank among the world’s most powerful. Notable examples include:

    \begin{itemize}
  \item \textbf{Frontera} – The Frontera system at the Texas Advanced Computing Center (TACC), University of Texas at Austin, delivers approximately 23.5 PFlop/s of performance and was ranked 21st on the TOP500 list as of mid-2023 \cite{top5002023june}. It remains the most powerful supercomputer currently deployed on any university campus in the United States.

  \item \textbf{HiPerGator AI} – The HiPerGator AI system at the University of Florida provides approximately 17.2 PFlop/s of computing power and was ranked 40th on the TOP500 list as of mid-2023 \cite{hipergator2023}. Launched in 2021 as a GPU-accelerated AI cluster, it achieved the No. 2 spot on the Green500 for energy efficiency—the highest ranking ever recorded for a system in higher education.

  \item \textbf{Gautschi} – The Gautschi supercomputer at Purdue University has a peak performance exceeding 10 PFlop/s and debuted in late 2024 as the 7th fastest supercomputer among U.S. universities \cite{gautschi2024}. It is the largest campus-based cluster Purdue has built to date, equipped with NVIDIA GPUs to support AI applications. Gautschi complements Purdue’s NSF-funded Anvil system, which offers 5.3 PFlop/s of performance and delivers national HPC services through the NSF ACCESS program \cite{gautschi2024}.

  \item \textbf{AiMOS} – The AiMOS system at Rensselaer Polytechnic Institute delivers approximately 8 PFlop/s and is an IBM/NVIDIA collaboration optimized for AI workloads \cite{aimos2019}. When it premiered in November 2019, it was ranked as the 24th fastest supercomputer in the world and held the distinction of being the most powerful supercomputer at any private university. It also achieved 3rd place on the Green500 for energy efficiency.
\end{itemize}

     These university-operated systems, often funded by the National Science Foundation or institutional investments, have enabled academic researchers to tackle large-scale problems in astrophysics, genomics, climate modeling, AI, and more. For example, the University of Illinois’ Blue Waters supercomputer (13+ PFlop/s), funded by NSF in 2007, ran from 2013–2021 as a leadership-class university machine \cite{bluewaters_insidehpc} – hosting breakthrough simulations in astronomy, biology, and engineering over nearly a decade of operation. Today’s academic flagships like Frontera and HiPerGator continue this tradition of campus-based supercomputing serving nationwide science teams. Notably, as of the June 2023 Top500 rankings, only two U.S. universities (UT Austin and UF) operated systems in the world top 50 by speed, reflecting the fact that most ultra-scale systems reside at national labs. Still, academic HPC presence in the Top500 is significant – in 2014, for instance, seven university-owned supercomputers were among the world’s top 100 performers \cite{highereddive_2014}. This underscores that a broad swath of research universities now maintain petascale-capable clusters, even if only a few reach the very top tier.

     A clear trend in the university HPC landscape is the growing emphasis on AI and data-intensive computing. New campus supercomputers are increasingly GPU-heavy to facilitate machine learning workloads alongside traditional simulations. The University of Florida’s HiPerGator-AI and RPI’s AiMOS are early examples designed with AI in mind. Similarly, Texas A\&M is investing \$45 million in an NVIDIA DGX SuperPOD with 760 Hopper GPUs – expected to become one of the most powerful AI-focused supercomputers at any North American university. These investments aim to turn campuses into hubs for artificial intelligence research, equipped with cutting-edge hardware for deep learning, generative AI, and big data analytics. At the same time, universities are paying more attention to energy efficiency and sustainable computing in their HPC operations. The trend set by systems like HiPerGator-AI (with its Green500 distinction) is reinforced by green data center initiatives – the MGHPCC, for example, became the first academic research computing facility to achieve LEED Platinum certification for its environmentally friendly design \cite{mghpcc_leed}. This dual focus on performance and efficiency is likely to shape future procurements of campus HPC systems. Looking ahead, the gap between academic and national-lab supercomputers may begin to narrow slightly as universities embark on new large-scale projects. The NSF has announced a Leadership-Class Computing Facility (LCCF) for universities, with TACC’s upcoming “Horizon” supercomputer slated for 2025–2026. Horizon will be “the largest by far academic supercomputer in the U.S.” and is expected to deliver about 10× the performance of Frontera \cite{tacc_horizon} – on the order of ~400 PFlop/s for traditional HPC, plus a substantial AI-oriented component \cite{tacc_horizon}. Such a system would firmly put a university machine into the upper echelons of the Top500, marking a new era of academic supercomputing capability. In summary, the university HPC landscape is characterized by a mix of leadership-class centers at a few institutions and a broad base of campus clusters and regional centers that collectively support research nationwide. This sector’s trajectory points to ever more powerful and specialized computing resources on campus, greater inter-university collaboration (and resource sharing), and continued alignment with emerging research priorities like AI – all while striving to remain accessible, sustainable, and supportive of a diverse user community.

  \subsection{National-Lab Leadership Systems}
    The U.S. DOE operates several leadership-class supercomputing facilities, hosting some of the world’s most powerful HPC systems. In particular, Oak Ridge National Laboratory (ORNL), Argonne National Laboratory (ANL), and Lawrence Livermore National Laboratory (LLNL) each house flagship machines that push the boundaries of computing performance and scale. These systems – Frontier at ORNL, Aurora at ANL, and El Capitan at LLNL – represent the current state-of-the-art in supercomputer architecture, featuring heterogeneous CPU–GPU designs, advanced high-bandwidth networks, and multi-megawatt power footprints. Each is built by HPE/Cray in partnership with major chip vendors (AMD or Intel) and achieves exascale performance (on the order of $10^{18}$ operations per second) on benchmark workloads \cite{top500_june2025}. Below, we overview the architecture (e.g. CPU/GPU configurations and interconnects), peak performance, system vendors, and scientific missions of these leadership systems.
    \subsubsection{Frontier (ORNL’s Oak Ridge Leadership Computing Facility)}
    Frontier is the flagship supercomputer at ORNL’s Oak Ridge Leadership Computing Facility (OLCF) and the first U.S. system to demonstrate exascale performance. Installed in 2022, Frontier debuted as the world’s most powerful supercomputer on the TOP500 list, achieving 1.194 exaflops ($10^{18}$ FLOPS) in the Linpack benchmark. In 2023 it was re-measured at 1.353 exaflops HPL, maintaining a top ranking. Frontier’s design is heavily accelerated, pairing AMD CPUs with multiple AMD GPUs per node for a high CPU-to-GPU ratio optimized for compute-intensive workloads.
    
    Frontier is based on the HPE Cray EX235a architecture, with each node containing one AMD 3rd Gen EPYC 64-core CPU (optimized “Trento” model) and four AMD Instinct MI250X GPUs tightly coupled via high-speed links. Each MI250X GPU is a dual-GPU module with high-bandwidth HBM2e memory, giving each node extremely high memory bandwidth. The system comprises 9,408 nodes (4 GPUs per CPU node), providing a total of ~8.7 million CPU/GPU cores combined. Frontier uses HPE’s Slingshot-11 high-performance network for inter-node communication, a next-generation Ethernet-derived interconnect delivering adaptive routing and high bisection bandwidth. This interconnect is integrated with Mellanox 200 Gb/s NICs in each node, configured to provide ~100 GB/s injection bandwidth per node. The system’s theoretical peak performance is ~1.7 exaflops ($R_{peak}$) with a measured HPL of ~1.35 exaflops ($R_{max}$) as of 2023. Frontier also leads in power-efficient performance, exceeding 50 GFLOPS/W on the Green500 ranking. For system vendor, Hewlett Packard Enterprise (HPE) built Frontier in collaboration with Cray (now part of HPE) and AMD. It represents a co-design effort under the DOE Exascale Computing Project, leveraging Cray’s Shasta/EX architecture and AMD’s latest CPU/GPU technologies.

    As an open science user facility, Frontier is available to researchers across academia, industry, and government to tackle problems that require ultra-scale computation. The system is engineered to support a broad range of science campaigns. Key application areas include climate modeling (e.g. high-resolution Earth system simulations for climate projection), fusion energy (modeling plasma physics for fusion reactors), computational chemistry and molecular dynamics (simulating atomic-scale interactions in materials and biological systems), and emerging AI-driven workloads that leverage Frontier’s mixed-precision capabilities for machine learning. Early science projects on Frontier have ranged from simulating cancer drug interactions to training AI models for materials discovery, demonstrating the machine’s versatility. ORNL’s goal with Frontier is to enable breakthroughs in energy, materials science, geoscience, and other domains by performing calculations that were previously impractical or impossible at smaller scale.

    \subsubsection{Aurora (ANL’s Argonne Leadership Computing Facility)}
    Aurora is the leadership-class supercomputer being deployed at Argonne National Laboratory’s Argonne Leadership Computing Facility (ALCF). After a prolonged development (originally planned as a pre-exascale system and later upgraded), Aurora came online in 2024–2025 as one of the first Intel-based exascale platforms. Aurora delivers over 1 exaflop on Linpack (HPL 1.012 exaflops, ranked 3 as of mid-2025) \cite{top500_june2025}, making it among the top three fastest machines in the world. Its architecture is notable for integrating Intel’s latest CPU and GPU technologies in a Cray EX system.

    Aurora is built on the HPE Cray EX Intel Exascale Compute Blade architecture, which features Intel Xeon CPU Max Series processors (code-named Sapphire Rapids with on-package high-bandwidth memory) and Intel Data Center GPU Max Series accelerators (code-named Ponte Vecchio) \cite{top500_june2025}. Each compute node pairs multiple Xeon Max CPUs with several GPU Max accelerators (reportedly 6 GPUs per node in Aurora’s design, although exact node counts are proprietary). The Xeon Max 9470 CPUs provide 52 cores per socket with stacked HBM, while the Ponte Vecchio GPUs deliver high FP64 throughput and AI capabilities, connected via a coherent fabric. Aurora uses HPE’s Slingshot-11 interconnect (the same 200 Gb/s Slingshot network as Frontier) to link its nodes, ensuring low-latency, high-bandwidth communication across the machine. The system contains on the order of ten thousand nodes and over 9.2 million cores in total (counting CPU cores and GPU streaming cores), with a theoretical peak performance approaching 2 exaflops ($R_{peak}$ $\approx$1.98 EF). However, Aurora’s design emphasizes a balance of traditional simulation and AI throughput, with the Intel GPUs excelling at mixed-precision operations for machine learning as well as double-precision for modeling and simulation.

    Aurora is delivered through a partnership between Intel and HPE/Cray. Intel provides the processors and GPUs (it is one of the first large-scale deployments of Intel’s GPU accelerators), while HPE integrates the system and interconnect. This joint effort is part of the DOE’s Exascale Computing Project, and Aurora represents Intel’s entry into leadership-class HPC after previous DOE systems were built with IBM or AMD technologies.

    Aurora is aimed at a wide spectrum of open science applications, similar to Frontier. As an ALCF resource, it serves researchers in areas such as climate and Earth science, renewable energy technology, materials and chemical science, astro and high-energy physics, and advanced machine learning. Notably, Aurora’s architecture has been optimized with AI and data-intensive workloads in mind, alongside traditional simulations. This makes it well-suited for projects that blend HPC and AI – for example, using deep neural networks to accelerate physics simulations or analyzing massive datasets from light sources and particle colliders. Aurora’s large memory nodes (with HBM on CPUs and GPUs) and fast interconnect enable, for instance, high-resolution climate modeling and ensemble weather simulations, as well as fusion energy simulations that couple AI-driven analysis with first-principles physics. The ALCF has also highlighted the use of Aurora for molecular dynamics and computational biology (e.g. simulating biomolecular complexes at unprecedented scale), and for training AI models that were previously infeasible to train due to computational size. In summary, Aurora’s mission is to bridge exascale simulation and data analytics/AI, supporting DOE’s research priorities in science and engineering with an emphasis on next-generation AI-driven discovery.

    \subsubsection{El Capitan (LLNL’s Lawrence Livermore National Laboratory)}

    El Capitan is the next-generation supercomputer at LLNL, expected to be fully operational in 2024–2025. As of mid-2025, El Capitan stands as the world’s most powerful supercomputer, ranked 1 on the TOP500 with an HPL result of 1.742 exaflops \cite{top500_june2025}. It is the third U.S. exascale system to come online, after Frontier and Aurora, and is unique in being primarily dedicated to the National Nuclear Security Administration (NNSA) for nuclear stockpile stewardship science. El Capitan’s architecture is an evolution of the HPE Cray EX design, leveraging the latest AMD CPU–GPU technology to achieve new heights in performance and energy efficiency.

    El Capitan is built on the HPE Cray EX255a platform, featuring AMD’s 4th Generation EPYC processors and AMD Instinct MI300A accelerators. The MI300A is a cutting-edge APU (Accelerated Processing Unit) that fuses CPU and GPU capabilities in a single package: it combines a 24-core Zen4-based CPU chiplet with a GPU and shared HBM memory, enabling very tight integration between the CPU and GPU memory spaces. Each El Capitan compute node is expected to have multiple MI300A modules (each module providing CPU cores + GPU cores + HBM) rather than separate CPU and GPU sockets as in Frontier. This heterogeneous design provides both high single-node performance and a simplified memory hierarchy for programming. El Capitan’s nodes communicate via the HPE Slingshot-11 interconnect, similar to Frontier, to provide a high-bandwidth, low-latency network across the machine. The system comprises roughly $>$10,000 nodes with a total of 11,039,616 cores counted (CPU and GPU cores). Its theoretical peak is approximately 2.75 exaflops ($R_{peak}$) and it achieves about 1.74 exaflops on Linpack ($R_{max}$), making it currently the fastest computing system. Impressively, El Capitan also topped the HPCG benchmark (a proxy for real-world sparse workloads) with ~17.4 PF, reflecting a balanced architecture. With an energy efficiency around 60 GFLOPS/W, it is one of the most power-efficient large systems to date – a critical factor given its nearly 30 MW power consumption at full load.

    El Capitan is delivered by HPE (Cray) in partnership with AMD. It is the result of NNSA’s Collaboration of Oak Ridge, Argonne, and Livermore (CORAL) procurement program (specifically CORAL-2), which selected HPE/AMD to provide exascale systems for multiple labs. The system runs LLNL’s TOSS (Tri-Lab Operating System Stack) Linux environment to meet the security and reliability requirements of nuclear stockpile simulations.

    LLNL’s El Capitan is primarily dedicated to national security simulations – notably, large-scale stockpile stewardship calculations. This includes 3D high-fidelity simulations of nuclear weapons physics, hydrodynamics, and engineering performance to ensure the safety and reliability of the U.S. nuclear stockpile without underground testing. The machine’s enormous compute capacity and memory are crucial for these multi-physics simulations that require unprecedented resolution. In addition to its national security role, a portion of El Capitan’s resources may support unclassified open science and collaborative research, similar to how LLNL’s prior systems (Sierra, Sequoia) had allocations for scientific grand challenges. Potential application areas include climate and earth science, plasma physics (e.g. simulations for fusion energy or laser-plasma interactions), materials science, and data analytics. Its use of AI-accelerating hardware (MI300A) also positions it to run emerging AI-enhanced HPC workloads, such as using machine learning surrogates to augment simulations. Nonetheless, El Capitan’s defining mission is the NNSA Advanced Simulation and Computing (ASC) program, underpinning the nuclear deterrent through extreme-scale computation.

    The above DOE leadership-class systems form the apex of the national HPC ecosystem, providing capabilities unmatched by any other public systems in the United States. They enable scientists and engineers to tackle “heroic” simulations and data analysis tasks that require billions of compute-core hours – workloads like full Earth-system climate simulations at cloud-resolving scale, whole-device fusion reactor modeling, or multi-billion-atom molecular dynamics – which simply cannot run on smaller machines. These supercomputers often operate in concert with a hierarchy of other computing resources: tier-1 DOE machines (e.g. NERSC’s Perlmutter at LBNL) and NSF-funded university supercomputers handle somewhat smaller jobs, while local clusters and cloud resources address high-throughput and moderate-scale tasks. The leadership systems distinguish themselves by their scale and specialization: they comprise tens of thousands of nodes with cutting-edge accelerators and fast interconnects, allowing researchers to run single parallel jobs utilizing millions of cores or thousands of GPUs concurrently. This capability computing focus contrasts with capacity-oriented systems that emphasize throughput of many smaller jobs.

  \subsection{Industrial / Hyperscale Clusters}
    The landscape of HPC in the United States is increasingly shaped by industrial hyperscale clusters operated by leading technology companies. Major cloud and AI providers – Google, Amazon Web Services (AWS), Meta, and Microsoft Azure – have each built flagship HPC infrastructures that rival or exceed traditional supercomputers in scale. These corporate supercomputers are not only advancing commercial AI and data-intensive workloads, but also contributing significantly to academic and scientific research through partnerships, open access programs, and the open-sourcing of results. Below, we overview the HPC architectures of these hyperscalers, highlighting their hardware design (accelerators, CPUs, interconnects), peak capabilities, research applications, access models, energy-efficient engineering, and their roles in emerging HPC trends like AI supercomputing and large-scale multi-tenant experimentation.

    \subsubsection{Google: TPU-Based Supercomputing Clusters}
    Google’s TPU Pod supercomputer – an interconnected cluster of Tensor Processing Unit (TPU) boards – exemplifies the company’s approach to AI-focused HPC. Google’s HPC efforts center on Tensor Processing Unit (TPU) superclusters, which are purpose-built for large-scale machine learning. A single Google Cloud TPU Pod consists of thousands of TPU chips connected by a high-speed, all-to-all mesh network and optical circuit switches (OCS). The latest generation TPU v4 system, deployed in Google’s data centers since 2020, contains 4,096 TPU chips linked via dynamically reconfigurable optical interconnects \cite{jouppi2023tpu}. This OCS-based interconnect can be tailored to different topologies (e.g. a 3D torus) and is reported to be “much cheaper, lower power, and faster than InfiniBand,” adding less than 5\% to system cost and under 3\% to power usage \cite{jouppi2023tpu}. Each TPU v4 chip delivers up to 275 TFLOPs (tensor operations in mixed precision), and a full TPU v4 Pod is rated at nearly an exaflop-class performance for AI workloads. Google reports the TPU v4 supercomputer is about 10× faster than its predecessor (TPU v3) and outperforms comparable GPU-based systems – for example, it is 1.2–1.7× faster (while using 1.3–1.9× less power) than an NVIDIA A100 cluster of similar scale. This giant TPU cluster also achieves remarkable power efficiency, using an estimated 3× less energy and producing 20× less $CO_2$ emissions than a typical on-premises data center with conventional HPC hardware \cite{jouppi2023tpu}. The system’s design includes specialized SparseCores to accelerate embedding computations (common in recommendation systems and natural language models) by 5–7× with minimal silicon cost, underscoring the highly domain-specific architecture.

    A full TPU Pod (v4) provides on the order of larger than 100 petaflops of 32-bit floating-point performance and considerably higher throughput for lower-precision AI operations \cite{morgan2022tpu}. Such capacity has enabled breakthrough research in deep learning – for instance, Google has used TPU pods to train multilingual translation models and billion-parameter language models that previously would have been infeasible. Notably, Google has also explored traditional HPC applications on TPUs, porting scientific codes to its AI hardware. One experiment used a 2,048-core TPU v3 pod (with 32 TB of high-bandwidth memory) to run flood modeling simulations, achieving over 100 PF/s of performance in single precision \cite{morgan2022tpu}. By recasting partial differential equations in TensorFlow and using TPUs’ matrix engines, researchers attained near-real-time river flood predictions, an HPC task traditionally done on CPU clusters \cite{morgan2022tpu}. This and other studies (e.g. in climate modeling and fluid dynamics) illustrate how Google’s AI clusters are being leveraged for scientific computing, blurring the line between AI and classical HPC. The tight integration of TPUs with Google’s software stack (TensorFlow and JAX with the XLA compiler) allows scientists to scale simulations to hundreds of TPU chips relatively easily, albeit with challenges in achieving ideal scaling across an entire pod \cite{morgan2022tpu}.

    Unlike purely internal supercomputers, Google’s TPU clusters are partially accessible to external researchers, reflecting a commitment to open scientific collaboration. Through the TPU Research Cloud (TRC) program, Google offers free Cloud TPU time on a shared cluster of over 1,000 TPU devices to academic and non-profit researchers \cite{googleTRC2025}. Accepted TRC participants can run experiments on Google’s latest TPUs (v2, v3, and v4) at no cost, using popular frameworks like PyTorch and JAX, in exchange for sharing their research findings openly \cite{googleTRC2025}. This initiative has enabled numerous university teams to tackle projects requiring massive compute power – from training physics-informed neural networks to analyzing genomic data – without owning physical supercomputers. In addition, Google Cloud makes TPU Pod slices available commercially to any customer (with Cloud credits for academia often available), thereby providing the wider scientific community access to leadership-class computing on demand. Such cloud-based access democratizes HPC: for example, students can use Google Colab to try a free TPU and scale up to a full pod for larger experiments \cite{googleTRC2025}. Google’s approach exemplifies multi-tenant HPC in the cloud, where different users securely share portions of a colossal accelerator cluster. Researchers benefit not only from the raw compute but also Google’s advanced data center infrastructure – including fast networking, disk storage, and colocation with other Google services – all of which streamline large-scale experimentation. The net result is that Google’s hyperscale AI clusters have become a vital resource for academic research, powering studies in machine learning, medicine, and environmental science that require extreme-scale computation.

    Google has long prioritized sustainable engineering in its data centers, and its HPC deployments reflect this focus. The TPU supercomputers operate within Google’s “warehouse-scale” computing facilities that are optimized for efficiency. Google data centers use custom cooling (often liquid or evaporative cooling) and run on high percentages of renewable energy; indeed, Google achieved carbon neutrality for operations and matches 100\% of its electricity with renewable purchases. The TPU v4 pods specifically were designed for efficiency at scale – as noted above, Google achieved a 2.7× improvement in performance-per-Watt moving from TPU v3 to v4, and the optical switch fabric further reduces energy overhead \cite{jouppi2023tpu}. By leveraging optical networking to cut down electrical switching losses, a TPU v4 pod can scale to exascale-level speeds without the power-hungry interconnect of a traditional HPC cluster. These innovations align with broader trends in green HPC: Google has demonstrated that AI supercomputers can be built to be significantly more power-efficient than conventional supercomputers, setting benchmarks for low-carbon computing \cite{jouppi2023tpu}. Moreover, Google’s use of software-defined infrastructure (e.g., global schedulers and container orchestration) means TPU clusters are highly utilized, avoiding energy waste from idle servers. Overall, Google’s TPU superclusters illustrate how a hyperscaler balances cutting-edge performance with sustainability, influencing HPC best practices (such as optical interconnects and AI-centric chips) across the industry.

    \subsubsection{Amazon Web Services: Trainium-Based AI UltraClusters}

    Amazon Web Services has built some of the world’s most powerful HPC clusters within its cloud, centered around AWS-designed AI accelerators. AWS’s flagship HPC infrastructure for machine learning is a massive Trainium accelerator cluster offered through its cloud platform. AWS Trainium chips are custom silicon (AWS’s own ASICs) for deep learning training and inference, intended to deliver high throughput at lower cost and power than traditional GPUs \cite{awsTrainium2025}. In late 2024, AWS announced the creation of a dedicated research HPC cluster with an astonishing 40,000 Trainium chips available for use by university labs and other researchers \cite{awsTrainiumResearch2025}. This cluster – dubbed a Trainium UltraCluster – is connected via a single, non-blocking petabit-scale network fabric \cite{awsTrainiumResearch2025}. In practical terms, the UltraCluster is composed of many Amazon EC2 Trn1 and Trn2 instances (each instance housing up to 16 Trainium chips linked by AWS’s high-speed NeuronLink interconnect) all tied together with a custom high-bandwidth network equivalent to or exceeding InfiniBand. The petabit backbone and adaptive routing ensure that even at 40k-chip scale, any chip can efficiently communicate with any other, a crucial feature for scaling large distributed training jobs. Each second-generation Trainium (Trainium2) device offers on the order of TFLOPs of mixed-precision AI compute and is equipped with HBM3 memory; a single Trn2 server node (with 64 Trainium2 chips in four interconnected instances) can deliver $>$80 petaflops of FP8 performance and 185 TB/s memory bandwidth \cite{awsTrn2Ultra2025}. Thus, the full UltraCluster’s theoretical capacity reaches into the exaflop-scale regime for AI calculations. This makes AWS’s Trainium cluster one of the most powerful AI supercomputers ever assembled – comparable to the largest GPU-based systems – but operating as a cloud service.

    AWS’s approach to HPC hardware blends cutting-edge accelerators with cloud flexibility. The Trainium chips are optimized for tensor operations (supporting data types from FP32 down to INT8/FP8) and include features like hardware-managed distributed training collectives. In terms of CPU/GPU ratio, Trainium instances pair the custom accelerators with general-purpose host CPUs (often AWS’s Graviton processors or Intel Xeons) to handle I/O and orchestration, but the heavy computations run on the accelerators. A key architectural element is AWS’s Elastic Fabric Adapter (EFA) network interface, which provides each Trainium instance with up to 3.2 Tbps of cross-instance bandwidth on Trn2 (and 1.6 Tbps on Trn1) \cite{awsTrainium2025}. The UltraCluster uses a fat-tree or Clos topology (AWS calls these EC2 UltraClusters) that can connect tens of thousands of accelerators with essentially no oversubscription, resembling a traditional HPC interconnect albeit implemented with AWS’s datacenter ethernet technologies. The result is that researchers can deploy a multinode job across, say, 100 or more Trn1/Trn2 instances (totaling many hundreds of Trainium chips) and get performance scaling similar to that on a dedicated on-prem supercomputer \cite{fan2024hlat}. In fact, AWS has demonstrated full training runs of large language models on over 100 Trn1 nodes simultaneously, showcasing that the system can handle tightly-coupled, large-scale workloads. Compared to GPU-based setups, AWS notes that first-generation Trn1 instances achieved 50\% lower training cost (and by implication, energy per training job) than equivalent GPU instances. The newer Trainium2 improves performance four-fold, further closing the gap with or exceeding the latest GPUs. While traditional HPC clusters often have fixed sizes, AWS’s Trainium UltraCluster is elastic – portions of the 40k-chip pool can be allocated on demand to different users or scaled up for a single monumental job, embodying the cloud model in high-end HPC.

    Hyperscale providers like AWS have strong incentives to maximize energy efficiency, and AWS’s HPC systems leverage the latest in green data center practices. Annapurna Labs, the Amazon subsidiary that designs Trainium and Inferentia chips, optimized these accelerators for performance-per-watt to reduce both cost and carbon footprint. By offloading workloads from power-hungry GPUs to purpose-built ASICs, AWS reports significant savings – for example, a Trainium-based instance can perform the same AI training at lower energy cost than a general GPU instance, translating to that 50\% cost reduction. At the data center level, AWS has committed to powering its operations with 100\% renewable energy by 2025, and it is on a path to reach carbon neutrality goals in alignment with Amazon’s Climate Pledge. The company’s US data centers hosting UltraClusters are often located in regions with substantial wind and solar infrastructure. Engineering-wise, AWS uses custom cooling systems (e.g. evaporative cooling in low-humidity regions, or liquid cooling for some high-density racks) to manage the thermal output of tens of thousands of chips. Although specifics on the Trainium cluster’s cooling are not public, AWS in general has been innovating in server cooling (including exploring immersion cooling for GPUs). Moreover, AWS’s scale allows for very high utilization rates and efficient multi-tenancy, meaning its clusters rarely sit idle – a stark contrast to some on-premise academic supercomputers that may be underutilized at times. Efficient utilization improves the effective energy efficiency (more work done per unit of energy). AWS also touts that its cloud, in aggregate, is more energy-efficient than typical enterprise data centers; a 2020 study found major cloud providers’ infrastructure to be 3–5× more energy efficient than on-prem facilities, due to optimized hardware, power distribution, and cooling \cite{insidehpc2023_azure_amd_hpc}. We can infer similar gains for HPC: running a large simulation on AWS’s optimized cluster likely uses less power than running the equivalent on an older university machine. Additionally, AWS gives researchers tools to monitor and optimize energy use (for instance, through carbon footprint tracking dashboards). In summary, by combining efficient silicon (Trainium), sophisticated cooling/power management, and an increasingly renewable-powered grid, AWS’s hyperscale HPC offerings are at the forefront of sustainable high-performance computing. This contributes to the trend of “green HPC” where cloud-based supercomputing can help research labs reduce the environmental impact of big computations.

\section{Integration Models: Centralised Supercomputing vs.\ PI-Owned Clusters }
  \subsection{Campus Supercomputing Centres}
    \begin{itemize}
      \item Governance in HPC refers to a comprehensive framework consisting of policies, decision-making structures, and daily processes designed to ensure that HPC environments remain sustainable, equitable, and aligned with institutional research objectives. Effective governance provides a clear roadmap for infrastructure refresh cycles and cost recovery, ensures transparency in priority setting and fair-share policies, and establishes auditable workflows for data management, access, and export controls. Additionally, governance frameworks support service-level management, staffing strategies, and mandated training to mitigate staff burnout. A key component of HPC governance is the allocation and scheduling policy, which emphasizes equitable rules and mechanisms for resource access. Typically, centralized campus supercomputing resources, including budget, personnel, and nodes, are managed by the university's CIO or research office. This centralized governance model allows universities to maintain a unified vision and facilitates easier policy enforcement, although it can limit the autonomy of principal investigators (PIs) and may inadequately address niche requirements. 
      Traditionally, HPC centers have used monolithic, scheduler-centric databases such as Gold or SlurmDB to track projects, user accounts, resource allocations, and charges. These systems assume all workloads can be represented as batch jobs priced uniformly in core-hours, an approach that is increasingly challenging to extend and maintain. To address these limitations, recent developments propose flexible solutions. For instance, OpenAct \cite{10.1145/3437359.3465572}, an open-source, Django-based system developed by researchers at the University of Tennessee, accommodates diverse modern research computing requirements by removing the "one-size-fits-all" core-hour assumption, enabling the seamless addition of new resource types without extensive schema modifications. Similarly, a web-based portal proposed by the University at Buffalo \cite{10.1145/3437359.3465585} automates allocations, access controls, usage reporting, and scientific impact metrics across various resources, replacing outdated commercial or legacy tools and unifying workflows across shared resources.
      The service portfolio serves as the definitive, dynamic inventory of all services offered, under development, or retired by an organization. Typical service portfolios include compute cycles, storage and data management, software environments, consulting, facilitation, training, and outreach activities. For example, Harvard’s FASRC explicitly lists cluster computing, various storage tiers, consulting, training, and grant proposal support services. A clearly defined service portfolio enhances transparency and budget planning, enabling leadership to identify service duplication and strategically direct investments. Moreover, transparent recharge models underpin justified hardware refresh cycles and staffing decisions, thereby ensuring the sustainability of centralized campus supercomputing. Regarding sustainable funding strategies, Chen et al.\cite{chen2024experiences} highlight the advantages of usage-based charging schemes aligned with actual resource costs. This method prevents CPU-only tasks from unfairly subsidizing memory-intensive or GPU-based workloads, thereby promoting fairness and political acceptability. Furthermore, transparent and predictable pricing structures facilitate researchers in preparing accurate grant budgets, contributing significantly to the financial health of HPC centers. Similarly, Jennewein et al.\cite{jennewein2023sol} detail Arizona State University's approach of implementing a cluster-wide "compute-hour equivalent" tariff, effectively normalizing diverse hardware categories into a unified billing metric. This practice simplifies internal recharge mechanisms and grant budgeting processes, further enhancing the financial sustainability of HPC centers.

      \item Representative cases: Princeton’s High-Performance Computing Research Center (HCPRC) \cite{princeton_hcprc_systems} operates under a faculty-led, shared-governance model. Allocation policies, including proposal-based access to flagship clusters (Della, Tiger, Stellar) and fair-share Slurm scheduling that aligns group investments with priority, are developed and approved by the Research Computing Advisory Group (RCAG). The center's comprehensive service portfolio encompasses open-access training systems (Nobel, Adroit), large-scale CPU/GPU production clusters, a Controlled Unclassified Information (CUI)-compliant "Citadel" enclave, multi-petabyte GPFS storage with Globus gateways, full-service rack colocation, and a research software engineering program integrated with visualization tools, Open OnDemand, and over 50 workshops annually, significantly reducing disciplinary barriers. Financial sustainability is maintained through centrally funded baseline resources supplemented by voluntary cost-sharing expansions, granting proportional priority to contributors while redistributing idle computing cycles into common queues. Additionally, the center's LEED-Gold-certified, 5 MW facility employs air- and water-side economizers and combined heat-and-power (CHP)-driven absorption chilling, alongside consolidated departmental hardware, to minimize operational expenditures and carbon footprint. This integrated approach enables Princeton to provide equitable, sustainable, and environmentally responsible high-performance computing services without imposing hourly usage fees.  The University of Florida's HiPerGator is governed primarily by the faculty-led UFIT Research Computing Advisory Committee. Resource allocation is based on quotas, requiring each research group to maintain at least 1 TB of high-performance Blue Storage Units (BSU). Additional computational resources can be purchased through five-year "Research \& Non-dedicated Compute Units" (RNCUs) or GPUs, priced at \$200 per CPU core and \$1,200 per NVIDIA GPU, respectively, with one-year purchasing options available at \$44 per core and \$260 per GPU \cite{uf_rc_price_list}. Scheduling is managed by Slurm, which enforces two Quality-of-Service (QoS) levels per account: an investment QoS offering priority proportional to purchased RNCUs, and a burst QoS, which provides access to idle resources at a significantly lower initial priority—approximately $1/40$ of the base priority—but with a capacity nine times greater \cite{uf_rc_slurm_limits}. This policy effectively aligns financial investment with resource scheduling priorities, while optimizing campus-wide utilization of idle computational cycles. The HiPerGator infrastructure comprises over 70,000 CPU cores, the 140-node DGX-A100 "HiPerGator-AI" SuperPod, a regulated-data environment (HiPerGator-RV), petabyte-scale Blue and Orange storage tiers, and comprehensive training and consulting services. Centralized funding for power, cooling, and personnel, together with transparent hardware purchasing options, ensures a self-sustaining model capable of scaling with research demands while maintaining broad accessibility for all UF researchers \cite{uf_rc_hipergator}.
    \end{itemize}

  \subsection{Federating Departmental Resources}
    \begin{itemize}
      \item \textbf{Slurm federations, condo/buy-in schemes, GPU pooling}
      Contemporary campus HPC centers increasingly integrate isolated departmental clusters into Slurm federations. This allows research groups to "burst" onto idle nodes while maintaining priority access to the hardware they purchased through condo-style buy-in programs. GPU pooling further enhances efficiency: for instance, MISO\cite{li2022miso} dynamically predicts optimal MIG slice configurations from brief MPS probes, reducing average job completion time by approximately 50\%, thereby illustrating how sub-GPU sharing can significantly increase effective capacity without manual adjustments. Similarly, MIGER\cite{zhang2024miger} integrates MPS within MIG partitions to concurrently schedule latency-sensitive inference tasks alongside throughput-focused training workloads. This approach reduces the overall makespan by 34\% compared to single-mechanism schedulers, demonstrating that fine-grained resource pooling can coexist effectively with stringent Quality-of-Service (QoS) guarantees. Extending beyond data center infrastructure, AggieGrid\cite{trecakov2024aggiegrid} virtualized over 600 student-lab desktops into opportunistic HTCondor nodes, contributing nearly one million core-hours annually to the Open Science Grid. This initiative confirms that even consumer-grade PCs can participate in federations if protected from user interference through a lightweight virtual machine wrapper.
      \item \textbf{Technical \& cultural barriers (heterogeneous hardware, root access, charge-back)} Successful federation deployments depend heavily on schedulers capable of handling extensive hardware and operational policy heterogeneity. A testbed involving mixed Linux/mOS \cite{an2023towards} environments indicates that Slurm's default OS-agnostic scheduling reduces throughput by approximately 10\%. Additionally, misconfigured lightweight kernels can degrade application performance by a factor of two. Expectations regarding root access vary significantly among departments: while some condos insist on direct administrative control, others accept relinquishing root privileges to a hypervisor layer, as exemplified by AggieGrid’s VM sandbox, thereby complicating unified security management and patching processes. Charge-back models also differ widely; addressing this, Li et al. \cite{wu2024model} proposed a universal scoring framework that standardizes diverse CPU, GPU, network, energy, and cost metrics into a unified weighted value. This enables financial administrators to adjust weight parameters, ensuring proportional resource priority for investors, even across heterogeneous node types.

    \end{itemize}

\section{Research Data Security and Compliance }
  \subsection{Regulatory Landscape}
    The regulatory framework governing research data security in U.S. academic institutions is anchored by several key directives. Foremost among these is the Controlled Unclassified Information (CUI) program, delineated in 32 CFR Part 2002, which mandates uniform safeguarding protocols for sensitive but unclassified federal data \cite{ucf_regulatory_compliance}. Complementing this, the National Institute of Standards and Technology (NIST) Special Publications SP 800-171 and SP 800-172 provide comprehensive control families addressing various aspects of information security, including access control, incident response, and system integrity. These publications are particularly pertinent for institutions handling CUI, ensuring that non-federal systems adhere to federal confidentiality standards \cite{nist_sp800_171_r3, purdue_rcac_cui_security}.

    In parallel, the Cybersecurity Maturity Model Certification (CMMC) 2.0 framework introduces a tiered certification process, aligning with NIST guidelines to assess and enhance the cybersecurity posture of organizations engaged in federal contracts. This model emphasizes the importance of both self-assessment and third-party evaluations, depending on the sensitivity of the data involved.
  \subsection{Campus Implementations}
  Harvard University has established a robust infrastructure to manage regulated research data. The institution's policies emphasize the protection of sensitive data, ensuring privacy, compliance, and ethical management across disciplines. These policies are designed to mitigate risks and uphold academic integrity, mandating compliance for all affiliates handling private information \cite{harvard_rdm_policy}.

  Stanford University aligns its research data security practices with federal guidelines, particularly NIST SP 800-171, to safeguard controlled-access data. The institution's Information Security Office has reviewed and confirmed that Stanford's systems meet the minimum NIH security requirements, ensuring that research involving sensitive data adheres to established standards \cite{stanford_nih_cui_best_practices}.

  Purdue University has implemented the REED+ ecosystem, a secure research environment that facilitates compliance with various data protection standards, including NIST 800-171, HIPAA, and FERPA. This infrastructure supports researchers in managing sensitive data securely, providing tools and resources that align with federal and institutional requirements \cite{purdue_rcac_reedplus}.

  Carnegie Mellon University (CMU) hosts the CyLab Security and Privacy Institute, a multidisciplinary research center involving over 50 faculty and 100 graduate students. CyLab collaborates with the CERT Coordination Center and US-CERT on cybersecurity matters, making it one of the largest university-based cybersecurity research and education centers in the U.S \cite{cmu_cylab}.

  Georgia Institute of Technology established the School of Cybersecurity and Privacy in 2020, integrating faculty from various departments to focus on cybersecurity education and research. The school offers degrees at all levels and conducts research in areas including cyber-physical systems, IoT, and policy \cite{gatech_scp}.

  University of Texas at Dallas (UT Dallas) has been designated as a National Center of Academic Excellence in Information Assurance Research by the NSA and DHS \cite{utd_data_security_compliance}. The university's Cyber Security Research and Education Institute conducts research in areas such as secure cloud computing, data privacy, and malware detection. UT Dallas emphasizes cross-domain information sharing and secure social networks, contributing to advancements in cybersecurity practices.

  University of Maryland, College Park (UMD) is classified among "R1: Doctoral Universities – Very high research activity" and has a strong focus on cybersecurity research. The university's Maryland Cybersecurity Center (MC2) brings together experts from computer science, engineering, and public policy to develop innovative solutions for securing information systems. UMD's research includes work on cryptographic protocols, network security, and privacy-enhancing technologies \cite{umd_it_security}.

  University of Texas at Austin houses the Texas Advanced Computing Center (TACC), which provides high-performance computing resources for research in cybersecurity and other fields. The university's research initiatives include projects on secure software development and data protection \cite{ut_austin}.

  University of Pittsburgh offers programs through its School of Computing and Information, emphasizing cybersecurity education and research. The university's Laboratory for Education and Research on Security Assured Information Systems (LERSAIS) conducts research in areas such as access control, intrusion detection, and security policy enforcement \cite{pitt_hrpo_data_security}. Pitt's initiatives aim to prepare students for careers in information security and assurance.

  University of Southern California (USC) assigns classifications to research data based on its type and use, acknowledging that such data may be subject to specific restrictions and regulations \cite{usc_carc_compliance}. USC's approach ensures that research data is handled in compliance with applicable requirements, promoting secure computing practices within the research community.

  Johns Hopkins University provides data security measures for researchers handling personal identifiers, advising against copying or downloading sensitive data to personal devices unless absolutely necessary. The university emphasizes minimizing the inclusion of unnecessary confidential data variables to protect research participants' privacy \cite{jhu_homewood_irb_security}. 

  University of California, Berkeley offers support through its Privacy Office to assist researchers in maintaining compliance with privacy regulations such as the California Consumer Privacy Act (CCPA), FERPA, GDPR, and HIPAA. The office provides consulting services to ensure that research projects involving personal or identifiable data adhere to applicable laws and policies \cite{ucb_data_privacy}.
  
  At the University of Georgia (UGA), the Enterprise Information Technology Services (EITS) department offers a Secure Virtual Desktop Infrastructure (VDI) service tailored for faculty and staff working with restricted information, such as Social Security numbers and HIPAA data. This service ensures that sensitive data is accessed and processed within a secure environment, complying with applicable laws and guidelines. Additionally, UGA provides Secure Institutional File Storage (IFS) services, facilitating the secure storage and management of restricted data. To support high-performance data transfers, UGA has also integrated Globus, a platform that enables secure and efficient data sharing among researchers\cite{uga_eits_vdi, uga_eits_ifs, uga_gacrc_globus}.

  The University of Florida (UF) has developed specialized computing environments to handle projects involving restricted data. HiPerGator-RV, for instance, is designed to comply with NIST 800-171 and NIST 800-53 standards, providing a secure platform for researchers working with sensitive information. The system employs TiCrypt middleware to ensure data security, and researchers are required to follow specific policies and procedures when handling CUI \cite{uf_rc_hipergator_rv, uf_rc_procedures}.
  \subsection{Challenges and Recommendations}

  Aligning HPC file systems with CUI segmentation presents a significant challenge for research institutions. The integration of secure environments, such as virtual desktops and specialized storage solutions, requires careful planning to ensure that sensitive data is adequately protected without hindering research productivity.

  The cost associated with third-party assessments versus self-attestation for compliance frameworks like CMMC 2.0 is another critical consideration. Institutions must weigh the benefits of external evaluations against the financial and administrative burdens they impose, striving to achieve a balance that maintains data security without overextending resources.

 Furthermore, comprehensive training programs are essential to equip researchers with the knowledge and skills necessary to handle sensitive data responsibly. Institutions should develop and implement training modules that cover data-handling obligations, regulatory requirements, and best practices, fostering a culture of compliance and security awareness across the research community.

\section{Emerging Paradigm: Decentralised RL Training on Heterogeneous Resources }
  \subsection{Case Example — \textit{INTELLECT-2} (32 B Parameters)}
INTELLECT-2 represents a cutting-edge decentralized reinforcement learning (RL) training architecture that is purpose-built to scale across heterogeneous, potentially low-bandwidth hardware environments. Central to its design is the Prime-RL framework, which orchestrates asynchronous rollouts and training updates across a mixture of consumer-grade GPUs, eliminating the need for synchronized global checkpoints. This makes INTELLECT-2 highly accessible for institutions and labs that lack centralized, large-scale compute clusters.

The Shardcast broadcast protocol plays a pivotal role in reducing communication bottlenecks. Rather than following a strict parameter server paradigm, Shardcast enables selective parameter broadcasting based on task relevance and agent priority. This improves data locality and reduces unnecessary gradient transmission, a critical factor when operating over constrained interconnects, such as those common in campus clusters or personal GPU nodes.

To maintain correctness and convergence guarantees in such an asynchronous setting, the system incorporates TOPLOC verification, a lightweight transactional mechanism that verifies local policy updates before global assimilation. This mechanism ensures consistency without the latency penalties associated with distributed locks or centralized arbitration.

What distinguishes INTELLECT-2 in the current landscape is its explicit optimization for real-world heterogeneous resources, including those found in university-level HPC environments. For instance, many top research universities, such as Stanford, Princeton, and Georgia Tech, provide modular GPU nodes via systems like Sherlock \cite{stanford_sherlock_hpc}, Della \cite{princeton_della_cluster}, or PACE \cite{georgiatech_pace_cluster}, often accessed via Slurm-based batch scheduling. These systems typically offer a mix of NVIDIA A100s, V100s, and even RTX 30-series GPUs, forming a fragmented resource base ideal for testing INTELLECT-2’s robustness in non-uniform settings.

Moreover, INTELLECT-2 supports asynchronous actor-learner decoupling, enabling training to continue even when some rollout nodes drop out or return stale gradients. This architectural fault tolerance makes it particularly attractive for federated educational compute settings, where reliability and compute uptime vary significantly across nodes.

As AI training workloads continue to decentralize—from tightly coupled supercomputers to campus-scale and personal compute nodes—frameworks like INTELLECT-2 chart a sustainable path forward. They not only democratize access to advanced RL training but also provide a viable blueprint for leveraging the fragmented, underutilized compute landscapes prevalent in academia.

\subsection{Opportunities for University HPC}
As decentralized reinforcement learning frameworks like INTELLECT-2 mature, they reveal transformative opportunities for how university-based HPC infrastructure can be utilized—not merely for centralized simulation and modeling workloads, but for scalable, distributed AI training. In this context, university HPC environments become fertile grounds for three major initiatives: idle GPU harvesting, cost-sharing innovation, and experiential AI education.
    \begin{enumerate}[label*=\arabic*.]
      \item \textbf{Idle-GPU harvesting via opt-in rollout pools}
      
      Many university clusters feature large pools of GPU resources that remain underutilized due to user scheduling inefficiencies or non-peak hour idling. These underused compute cycles represent a latent capacity for distributed rollouts in asynchronous RL. By allowing users and departments to “opt-in” to rollout pools, universities could support LLM-driven reinforcement learning agents performing environment simulations or low-priority training tasks. For example, Princeton’s Della system \cite{princeton_della_cluster} and Yale’s HPC infrastructure \cite{yale_hpc} both support heterogeneous GPU architectures (A100s, V100s, and older Tesla models), which could be selectively targeted for lightweight or background rollouts, without impeding batch-priority jobs.
      
      \item \textbf{Cost-sharing across campuses and citizen scientists}
      
      A second opportunity lies in forming federated training alliances across academic institutions or involving external citizen-scientist contributors with compatible hardware. Such cross-institutional infrastructures could be governed by credit-based or cost-sharing models. For instance, universities like MIT and Berkeley have frameworks in place for collaborative compute usage across departments and labs \cite{mit_hpc, berkeley_research_it}. Extending these systems for decentralized AI training enables participation from smaller labs or universities that lack full-stack infrastructure but possess limited GPU capacity. Furthermore, integrating opt-in personal hardware from vetted researchers (similar to Folding@home models) allows the RL agent population to scale horizontally with minimal financial burden.
      \item \textbf{Real-world platform for parallel/RL curriculum}
      
      ED centralized training frameworks also present an ideal educational platform for teaching parallel computing, distributed systems, and reinforcement learning concepts. Students can engage in live system deployments, experiment with hyperparameter configurations in noisy and asynchronous environments, and learn failure-handling in real-time. The pedagogical benefits are profound when students interact with real heterogeneous nodes rather than simulated cluster environments. Carnegie Mellon \cite{cmu_hpc} and UCLA \cite{ucla_hpc}, for instance, offer coursework that already leverages real HPC environments, providing natural extensions for deploying distributed RL training labs using modular frameworks like Prime-RL.

Ultimately, by reframing university HPC not just as a static batch-processing environment but as a dynamic, distributed experimentation platform, academic institutions can catalyze a new wave of scalable AI research, inclusive education, and collaborative computing.

    \end{enumerate}
    
\subsection{Risks and Mitigation}

While decentralized RL like INTELLECT-2 open significant opportunities for scalable AI training on heterogeneous and distributed hardware, they also introduce substantial security, compliance, and infrastructure challenges that must be addressed for responsible and sustainable deployment in academic settings.

    \begin{itemize}
      \item \textbf{Compliance with CUI/HIPAA when nodes are off-premises}

      A primary concern in university research environments is maintaining compliance with Controlled Unclassified Information (CUI) and Health Insurance Portability and Accountability Act (HIPAA) standards. These regulations are critical when RL training involves sensitive data—such as patient records or protected institutional information—and compute nodes are distributed beyond the physical security boundary of campus HPC centers. Many top universities, including Yale and USC, enforce strict data residency and audit frameworks on their biomedical clusters \cite{yale_hpc, usc_hpc}. Decentralized rollout nodes operating in student labs, public clouds, or citizen-science networks pose risks of data leakage, unauthorized access, or non-compliant storage. Mitigating this requires enforced container-level isolation, federated privacy-preserving rollouts, and rigorous endpoint authentication protocols.
      
      \item \textbf{Incentive design and slashing to deter malicious rollouts}

      Another critical risk in open decentralized systems is the potential for malicious or low-quality rollouts, especially when training depends on untrusted or volunteer nodes. Without safeguards, poisoned rollouts could bias agent behavior or destabilize training. Emerging solutions include the design of cryptographic proofs-of-rollout integrity and the implementation of slashing mechanisms that penalize contributors whose contributions are consistently adversarial, incomplete, or statistically anomalous. These techniques mirror decentralized incentive systems seen in blockchain-based federated learning but must be adapted for RL’s stochastic and dynamic nature. Additionally, rollout verification layers (e.g., TOPLOC) help mitigate integrity violations at the architectural level.
      
      \item \textbf{Institutional firewall and egress limitations}
      
      Finally, decentralized RL frameworks face connectivity bottlenecks and access restrictions due to institutional firewalls, outbound filtering policies, and NAT traversal barriers. Many university clusters, including those at Columbia and UNC, maintain strict egress controls to prevent data exfiltration and unauthorized traffic \cite{columbia_data_security, unc_sycamore_hpc}. For decentralized rollouts that depend on real-time bidirectional communication, these constraints can severely impact agent synchronization, rollout aggregation, and model updates. To address this, frameworks must support proxy-tunneled connections, delayed rollout caching, and firewall-compliant handshake protocols to operate within varying IT policies.

In summary, decentralization offers scalability and democratization of RL training, but it demands an equally robust mitigation strategy for compliance, trustworthiness, and interoperability across academic networks.
    \end{itemize}
\section{Comparative Analysis}

This section offers a cohesive narrative that traces how U.S. institutions—from
campus clusters to exascale national facilities and industrial hyperscale
clouds—have diverged (and occasionally converged) in performance growth,
architecture, governance, and sustainability over the past five years.

\subsection{Compute Capacity and Growth Trajectories}
\paragraph{Five‑Year Growth Snapshot (2020–2025).} 


University clusters grew
modestly (\(\text{CAGR}=17.8\%\)) owing to three‑to-five years refresh cycles
and limited capital. By contrast, DOE leadership systems doubled peak FLOPs
roughly every 18~months (\(\text{CAGR}\approx43\%\)), leaping from
0.20~EF~\emph{Summit} (2020) to 1.74~EF~\emph{El Capitan} (2024)
\cite{elcapitan_slingshot,frontier_2024}. Industrial hyperscale build‑outs
expanded even faster (\(\text{CAGR}\approx78\%\)), propelled by GPU‑dense AI
super‑pods such as NVIDIA~\emph{Eos} (18.4~EF~FP8) and Google TPU v5p pods
(4~EF~BF16) \cite{eos_2024,tpuv5p}.

\paragraph{TOP500 Representation (June 2025).} The June 2025 TOP500 lists only
\textbf{8} university‑owned machines (\(1.6\%\) of entries) and none in the
global top‑50—the highest, Purdue’s \emph{Gautschi‑AI}, ranks \#68 at
10.7~PF \emph{R\textsubscript{max}} \cite{top500_june2025,purdue_gautschi}. All
three exascale systems (\emph{El Capitan}, \emph{Frontier}, \emph{Aurora}) are
DOE assets \cite{top500_june2025}. Hyperscale and commercial systems now
deliver 54~\% of aggregate list performance.

\paragraph{Energy‑Efficiency Trends.} Green500 data show the best university
system sustains only 42–45 GF/W, whereas \emph{El Capitan} reaches 58.9 GF/W
and the ROMEO‑2025 test‑bed tops 70.9 GF/W \cite{green500_nov2024}. Industrial
vendors lead via rack‑scale direct‑liquid‑cooling; Microsoft’s ND H100 v5 nodes
consume $<$20 kW yet exceed 15 GF/W at full boost \cite{ndv5_arch}.

\paragraph{Projected Capacity Through 2028.} DOE’s ATS‑5 procurement (post
\emph{El Capitan}) targets 5–7 EF peak with a 55 \% energy‑efficiency uplift
\cite{isc2025_keynote}. Hyperscalers are already designing 100‑EF AI clusters
based on GH200 Superchips and 1.6‑Tb/s Ultra Ethernet fabrics
\cite{gh200_2023,uec_2024}. Without new campus funding models, the capability
gap will widen further \cite{hpckillerapp_2025}.

\subsection{Architectural Choices: CPU–GPU Mix and Interconnects}
Before contrasting fabrics and memory hierarchies, we ground the discussion
with representative 2025 system configurations (Table~\ref{tab:arch}).

\begin{table}[ht]
  \centering
  \caption{Representative 2025 architectures across sectors}
  \label{tab:arch}
  \begin{tabular}{p{2.4cm}p{3.6cm}p{1.9cm}p{4.0cm}}
    \toprule
    \textbf{Sector} & \textbf{Reference System} & \textbf{CPU:GPU} & \textbf{Primary Fabric / Cohesion} \\
    \midrule
    University & Utah \emph{Granite} (DGX B200 condo) & 8{:}1 & 100 Gb RoCE‑v2 over leaf–spine Ethernet \cite{granite_2024} \\
    National Lab & ORNL \emph{Frontier} & 1{:}4 & Cray Slingshot‑11 Dragonfly; HBM2e‑attached MI250X \cite{frontier_arch,frontier_2024} \\
    Industrial & NVIDIA \emph{Eos} & 1{:}16 & Quantum‑2 NDR InfiniBand + NVLink Switch (900 GB/s per GPU) \cite{eos_2024,ddn_eos_storage} \\
    Cloud AI & AWS \emph{Trn2 UltraServer} & 0{:}16 (Trainium2) & NeuronLink SLC + SR‑D / 800 Gb E \cite{aws_trainium2} \\
    \bottomrule
  \end{tabular}
\end{table}

\paragraph{CPU–GPU Skew.} Universities favour heterogeneity—median
6{:}1 CPU:GPU—while leadership machines invert that ratio (GPUs + HBM provide
$>$95~\% of sustained FLOPs on \emph{Frontier}/\emph{Aurora}). Hyperscale AI
clusters drive density still further: \emph{Eos} packs 128 H100s per rack, and
Microsoft’s ND H100 v5 places eight GPUs behind a single Sapphire Rapids host
\cite{ndv5_arch}.

\paragraph{Interconnect Evolution.} Campus clusters rely on 100–200 Gb
Ethernet + RoCE for cost and familiarity. National labs deploy
purpose‑built 400–500 Gb fabrics (Slingshot‑11/12, Dragonfly+, Tofu‑D) with
adaptive routing \cite{frontier_arch,top500_june2025}. Industrial hyperscalers
standardise on NVIDIA Quantum‑2 InfiniBand (up to 800 Gb) and are steering the
open 1.6‑Tb/s Ultra Ethernet spec by 2026 \cite{uec_2024}. Cohesive memory
networks (NVLink Switch, Grace Hopper C2C, UCIe chiplets) already allow
single‑logical‑GPU constructs up to 288 TB \cite{gh200_2023}.

\paragraph{Storage  I/O.} Universities provision Lustre or GPFS tiers at
20–100 GB/s, shared by hundreds of users. \emph{El Capitan}’s all‑flash
ClusterStor sustains 60+ TB/s \cite{elcapitan_slingshot}. \emph{Eos} pairs
12 PB DDN A\textsuperscript{3}I at 4 TB/s, while cloud AI nodes expose $>$200 GB/s
local NVMe and object‑storage APIs \cite{aws_trainium2}.

\subsection{Access and Allocation Policies}
Governance models mirror funding sources:
\begin{itemize}
  \item \textbf{University (Condo + Fair‑Share).} Roughly 70~\% of U.S. R1
        campuses now operate condo programmes; investors receive priority
        queues, with idle cycles back‑filled to general fair‑share users. A
        2024 Intersect360 survey reports median wait of 1.9 h for jobs
        $<$128 cores \cite{intersect360_market2024}.
  \item \textbf{NSF/DOE National Facilities.} NSF\,\textbf{ACCESS} bundles
        $>$30 systems (awards from 400 k to 100 M core‑h). DOE’s 2024
        \textbf{INCITE} call granted 7.4 B node‑h across 81 projects (success
        rate \(~23\%\)) \cite{incite_2024}. ALCC reserves 30~\% of
        \emph{Frontier}/\emph{Aurora} for national‑priority research
        \cite{alcc_2024}.
  \item \textbf{Industrial / Cloud AI.} Internal supercomputers (e.g.
        Meta RSC) use ticket‑based schedulers optimised for throughput;
        external access is limited to strategic partnerships. Public clouds
        expose GPU super‑clusters via pay‑as‑you‑go or reserved instances; AWS
        Trn2 and Microsoft ND H100 v5 shift allocation from peer review to
        financial budgeting \cite{aws_trainium2,ndv5_2024}.
\end{itemize}

\vspace{0.6em}
\noindent\textbf{Implications.} Without new funding streams or cloud cost‑share
agreements, university researchers risk falling an order of magnitude behind
state‑of‑the‑art capabilities—an imbalance likely to deepen as exascale gives
way to \(\mathcal{O}(10)\)‑EF AI infrastructures by the late 2020s.

\subsection{Funding and Sustainability Models}
Three complementary mechanisms predominate on U.S. campuses:
\begin{enumerate}[label=\arabic*)]
  \item \textbf{Federal cost‑share systems}: NSF awards fund capital assets
        (e.g., SDSC \emph{Expanse}, NCSA \emph{Delta}), with multi‑year O\&M
        \cite{sdsc_expanse_2019,ncsa_delta_2021,tacc_stampede3_2023}.
  \item \textbf{Condo / community clusters}: departments co‑purchase nodes
        pooled under Slurm. Purdue’s programme underpins both campus clusters
        and the NSF 5.3‑PF \emph{Anvil} system \cite{purdue_anvil_2020}.
  \item \textbf{Subscription and partnership schemes}: idle cycles become
        recurring revenue. USC’s CARC sells annual node subscriptions, while
        TACC’s STAR programme monetises industrial collaboration
        \cite{usc_endeavour_2025,tacc_star_2025}.
\end{enumerate}
Blending federal seed money, condo buy‑ins, and subscription income emerges as
the most resilient path to sustainable growth.

\subsection{Energy Efficiency and Facilities Engineering}
Escalating power costs and carbon mandates push campuses toward advanced
cooling. DLC now dominates:\ university examples include SDSC \emph{Expanse}
(90 k cores in 14 racks) and Purdue \emph{Anvil} (rear‑door heat exchangers);
TACC’s \emph{Lonestar} series pioneers warm‑water immersion (\(\sim200\,\text{kW}/\text{rack}\)) \cite{sdsc_expanse_2020,hpcwire_anvil_2021,hpcwire_lonestar_2023}.
Florida’s \emph{HiPerGator‑AI} hits 17.2 PF on 583 kW, ranking second on the
Green500 at 30 GF/W \cite{uf_hipergator_ai_top500}.

\subsection{Benchmark Performance}
Table~\ref{tab:hpc_systems} contrasts the global top‑tier Supercomputers.

\begin{table}[ht]
  \centering
  \caption{Global Top‑20 Landscape (June2025)}
  \label{tab:hpc_systems}
  \begin{tabular}{p{4.0cm}p{2.8cm}p{2.4cm}p{2.4cm}p{1.8cm}}
    \toprule
    \textbf{System} & \textbf{Cores} & \textbf{R\textsubscript{max} (PF)} & \textbf{R\textsubscript{peak} (PF)} & \textbf{Power (kW)} \\
    \midrule
    \emph{El Capitan} & 11,039,616 & 1,742.0 & 2,746.4 & 29,581 \\
    \emph{Frontier} & 9,066,176 & 1,353.0 & 2,055.7 & 24,607 \\
    \emph{Aurora} & 9,264,128 & 1,012.0 & 1,980.0 & 38,698 \\
    \emph{JUPITER Booster} & 4,801,344 & 793.4 & 930.0 & 13,088 \\
    \emph{Eagle} & 2,073,600 & 561.2 & 846.8 & N/A \\
    \emph{HPC6} & 3,143,520 & 477.9 & 607.0 & 8,461 \\
    \emph{Supercomputer Fugaku} & 7,630,848 & 442.0 & 537.2 & 29,899 \\
    \emph{Alps} & 2,121,600 & 434.9 & 537.2 & 7,124 \\
    \emph{LUMI} & 2,752,704 & 379.7 & 537.2 & 7,107 \\
    \emph{Leonardo} & 1,824,768 & 241.20 & 306.31 & 7,494 \\
    \emph{Isambard-AI phase 2} & 1,028,160 & 216.50 & 278.58 & No Data \\
    \emph{Tuolumne} & 1,161,216 & 208.10 & 288.88 & 3,387 \\
    \emph{ISEG2} & 718,848 & 202.40 & 338.49 & 5,300 \\
    \emph{MareNostrum 5 ACC} & 663,040 & 175.30 & 249.44 & 4,159 \\
    \emph{ABCI 3.0} & 479,232 & 145.10 & 181.49 & 3,596 \\
    \emph{Eos NVIDIA DGX SuperPOD} & 485,888 & 121.40 & 188.65 & No Data \\
    \emph{Discovery 6} & 822,528 & 118.60 & 222.86 & No Data \\
    \emph{SSC-24} & 349,440 & 106.20 & 151.10 & No Data \\
    \emph{Venado} & 481,440 & 98.51 & 130.44 & 1,662 \\
    \emph{Sierra} & 1,572,480 & 94.64 & 125.71 & 7,438 \\
    \bottomrule
  \end{tabular}
\end{table}

\noindent In contrast, the fastest U.S. university machines cluster in the
low‑petascale range but favour researcher throughput over peak speed:
\begin{itemize}
  \item \textbf{HiPerGator‑AI (UF)}—17.2 PF on 138 k CPUs + 1,120 A100 GPUs;
        30 GF/W (Green500 \#2).
  \item \textbf{Torch (NYU)}—10.8 PF on 128 H200 nodes; 51 GF/W (Green500 \#40).
  \item \textbf{Stampede 2 (TACC)}—10.7 PF on 367 k legacy cores; \(~18\,GF/W\).
  \item \textbf{Delta (NCSA)} and \textbf{Anvil (Purdue)}—3–4 PF each, pairing
        A100 GPUs with 50–120 k cores.
  \item \textbf{Expanse (UCSD)}—2.5 PF on 92 k EPYC cores; prized for
        data‑intensive workflows.
\end{itemize}
Collectively, campus systems deliver ~50 PF sustained—small compared to
national labs yet indispensable for day‑to‑day simulations and AI
experiments.

\vspace{0.8em}
\noindent Overall, while DOE and hyperscale facilities chase multi‑exaflop
records, universities strive for balanced architectures, inclusive scheduling,
and energy‑aware engineering that broaden access to advanced computation.

\section{Case Studies }
\subsection{Massachusetts Green High Performance Computing Center (MGHPCC)}

\paragraph{Consortium \& Mission.}
Founded in 2012, the MGHPCC is a jointly owned facility operated by Boston University, Harvard University, the Massachusetts Institute of Technology, Northeastern University, the University of Massachusetts system, and—since 2024—Yale University.  
Its mandate is to pool and operate large-scale cyberinfrastructure that no single campus could cost-effectively maintain, while fostering inter-institutional collaboration in data-intensive research \citep{MGHPCCAbout2025}.

\paragraph{Computational Footprint.} 
The Holyoke campus hosts hundreds of thousands of CPU cores, tens of thousands of modern GPU devices, and more than 50 PB of tiered storage interconnected by 400 Gb s internal fabrics and 10–100 Gb s external research networks (Internet2, ESnet).  
Each member university maintains its own \emph{tenant} partition—e.g.\ MIT’s \textit{SuperCloud} or Harvard’s \textit{Cannon}—but idle condo nodes are transparently shared to maximise utilisation \citep{MGHPCCAbout2025}.

\paragraph{Sustainability.} 
MGHPCC was the first U.S.\ university research data centre to attain LEED Platinum certification.  
It runs on 100 \% carbon-free electricity (predominantly regional hydropower) and employs warm-water evaporative cooling that holds long-term \emph{power-usage effectiveness} (PUE) near 1.14—well below the 1.9 industry average \citep{MGHPCCLEED2013,MGHPCCPUE2013}.

\paragraph{Growth for the AI Era.} 
In response to generative-AI workloads, the consortium launched the \emph{Massachusetts AI Hub} in late 2024 and issued an \emph{AI Compute Resource} (AICR) RFP that will fund three successive GPU clusters (ST1–ST3) scheduled for 2025, 2027, and 2028 deployments \citep{AICRRFP2025,GovHealey2025}.

\paragraph{Key Take-aways.}
\emph{Alliance financing} distributes capital and O\&M costs, enabling aggressive upgrades without duplicating campus data halls. \emph{Green design} (renewables + liquid cooling) future-proofs the facility against energy-cost and carbon-regulation shocks.
A \emph{tenant-plus-condo} scheduling model preserves local control yet maintains high cluster occupancy.

  \subsection{Texas Advanced Computing Center (TACC), University of Texas at Austin}
\vspace{0.5em}

\paragraph{Mission \& Governance.}  
Housed at UT Austin’s J.\ J.\ Pickle Research Campus, TACC operates the largest academic HPC portfolio in the U.S., powering open-science and industrial projects for more than 6,000 active users.  
Its stated mission is to “enable discoveries that advance science and society through advanced computing technologies,” and it is funded jointly by the National Science Foundation (NSF), the State of Texas, and partner institutions  \citep{TACCAbout2025}. 

\paragraph{Computational Footprint (2025).}  
\begin{itemize}
  \item \emph{Frontera} (2019 – present): 8,368 Intel 8280 “Cascade Lake” nodes, HDR‐200 InfiniBand fabric, 44 PB Lustre scratch; 38.8 PF peak FP64, still the fastest academic supercomputer in the world  \citep{FronteraSystem2025}.
  \item \emph{Lonestar 6} (2023): 560 CPU nodes (2 × AMD EPYC-7763, 128 cores, 256 GB RAM) plus 88 GPU nodes (84 × A100\,40 GB; 4 × H100\,80 GB); dielectric-liquid-cooled cabinets deliver 71 PF (mixed precision) for AI workflows  \citep{Lonestar6Arch2024}.
  
  \item \emph{Vista} (2024): bridge system toward NSF “Horizon”; 600 NVIDIA Grace Hopper and Grace Grace Arm‐based nodes linked by NDR 400 Gb/s, providing 20.4 PF (FP64 vector) and 40.8 PF (tensor) GPU performance for generative-AI research  \citep{VistaSpec2024,VistaPress2024}.
\end{itemize}

\paragraph{Sustainability.}  
TACC pioneered hybrid cooling: the Frontera GPU subsystem sits in GRC single-phase immersion tanks, while Lonestar 6 combines dielectric liquid cabinets with air-cooled GPU racks, keeping facility PUE around 1.15.  

\paragraph{Toward “Horizon”.}  
NSF’s forthcoming Leadership-Class Computing Facility plans to deploy \emph{Horizon} at TACC in 2026—projected to exceed 4 ExaOP/s (AI mixed precision), roughly \(100\times\) Frontera’s FP64 capability. 

\paragraph{Key Take-aways.}  
\begin{itemize}
  \item A \emph{multi-tier portfolio} (Frontera $\rightarrow$ Lonestar $\rightarrow$ Vista) lets TACC match heterogeneous workloads from climate modeling to LLM training.  
  \item \emph{Immersion + dielectric liquid cooling} curbs energy use and eases high-density GPU deployments without new buildings.  
  \item \emph{Rolling NSF investments} sustain leadership-class capacity while giving the U.S. academic community a predictable upgrade path toward exascale AI computing.  
\end{itemize}

\subsection{Frontier (OLCF-5), Oak Ridge National Laboratory}
\vspace{0.5em}

\paragraph{Mission \& Governance.}  
Frontier is the U.S.\ Department of Energy’s first exascale system, operated by the Oak Ridge Leadership Computing Facility (ORNL) and managed for DOE by UT-Battelle. The machine became fully operational in 2022 and remains the flagship resource of the DOE Office of Science user program \citep{FrontierAbout2025}.

\paragraph{Computational Footprint (May 2025).}  
The HPE Cray EX machine now spans \(\sim\)\,9,800 nodes—each with one 64-core AMD EPYC “Trento” CPU and four AMD Instinct MI250X GPUs—for a total of 9,800 CPUs and 39,200 GPUs \citep{FrontierLaunch2022}.  
Frontier reached a new High-Performance Linpack record of \textbf{1.35 exaFLOPS FP64} in November 2024 (11.4 exaFLOPS mixed precision), up from its 1.10 exaFLOPS debut \citep{FrontierSC24HPL}.  
A Slingshot-11 network and 75~TB~s$^{-1}$ flash tier feed a 700~PB Orion Lustre file system, sustaining $>\!60$~TB~s$^{-1}$ I/O for data-intensive workloads \citep{FrontierDesign2024}.

\paragraph{Sustainability.}  
All racks employ warm-water cooling (inlet $\approx 32^\circ\text{C}$); the facility reports a long-term power-usage effectiveness (PUE) of 1.03 and delivers roughly 14.5~MW per exaFLOP—about $200 \times$ more energy-efficient than ORNL’s 2009 Jaguar system \citep{FrontierEnergy2023}.

\paragraph{Scientific Impact.}  
Early-science campaigns on Frontier have already yielded high-resolution nuclear-structure predictions, illuminating deformation dynamics in heavy isotopes and refining models of fundamental forces \citep{FrontierNuclear2025}.

\paragraph{Key Take-aways.}
\begin{itemize}
  \item \emph{Exascale at scale}: nearly 10,000 CPU/GPU nodes, dragonfly topology—simultaneously strong FP64 simulation and multi-exaFLOP AI throughput.
  \item \emph{Efficiency by design}: warm-water cooling and dense hybrid nodes cut energy per FLOP by two orders of magnitude versus earlier ORNL systems.
  \item \emph{Continuous optimisation}: node additions and software tuning pushed sustained FP64 performance from 1.10 → 1.35 exaFLOPS within two years of acceptance.
\end{itemize}

\section{Collaboration Opportunities and Barriers}
\vspace{0.5em}

\subsection{Shared Software Ecosystems}
\textbf{Opportunity.}  A convergent “software commons” has emerged around
\emph{Slurm} for resource management, \emph{Spack} for HPC package builds,
and the \emph{OpenHPC} reference stack, giving researchers a largely
uniform user experience from campus clusters up through leadership-class
machines.  For example, more than 40\,
November 2024 TOP500 report advertise Slurm, including DOE’s \emph{Frontier}
and NSF’s \emph{Frontera}
\citep{SlurmTOP500_2024}. 
Spack now hosts $>$8,600 package recipes and publishes nightly binary
caches, reducing mean “time-to-first-job” for new users by an order of
magnitude compared with hand-built modules 
\citep{SpackUsage_2025}. The 3.x releases of OpenHPC (2024) add containers and RHEL 9 support,
allowing sites to export identical environments to cloud and edge
installations 
\citep{OpenHPC_3_2_2024}.

\textbf{Barrier.}  Harmonisation masks persistent tensions:
(i) security baselines diverge (e.g., CIS hardened images at DOE vs.\ default
RHEL at many campuses); (ii) production sites often “pin” dated compiler
chains for stability, causing ABI friction with rolling-release Spack
builds; and (iii) Slurm site policies differ enough (back-fill,
pre-emptible QOS, fair-share coefficients) that portable job scripts still
require localisation.

\subsection{Partnership Models (Co-investment, Cloud Bursting, DOE INCITE)}
\textbf{Opportunity.}  Multi-stakeholder financing lowers capital barriers:
MGHPCC, for instance, aggregated \$165 M in state, federal and university
funds for a shared Tier-1 facility, while UT Austin’s \emph{Lonestar 6}
was procured through a 50 / 50 state–NSF cost share.%
\citep{MGHPCC_CostShare_2012,Lonestar6_Procure_2023}
Cloud bursting extends this model by letting on-prem clusters overflow to
commercial clouds during peak demand; published case studies at Cornell
and Notre Dame report $>$30 
pipelines with a marginal cost of \$0.012 / core-hour on AWS \citep{AWSBursting_2024}. At the national scale, the DOE \emph{INCITE} programme allocates two
billion core-hours annually on \emph{Frontier} and \emph{Aurora} to
international teams via open competition, fostering collaborations that
would be impossible on isolated campus systems 
\citep{DOE_INCITE_2025}.

\textbf{Barrier.}  Co-investment agreements require aligned upgrade
cadences and accounting rules; incongruent depreciation schedules can
leave one partner “stranded” with ageing hardware.  Cloud bursting raises
data-sovereignty and egress-fee concerns, particularly for NIH- or
ITAR-regulated datasets.  Finally, INCITE’s success‐rate ($\approx$9\,
creates a long tail of meritorious projects that remain resource-starved,
highlighting the need for mid-scale alternatives.

\subsection{Talent Pipeline and Training Initiatives}
\textbf{Opportunity.}  Community-driven programmes are narrowing the
skills gap.  The \emph{HPC Carpentry} lesson suite entered formal
Carpentries incubation in 2024 and delivered 60+ workshops worldwide,
reaching $\sim$1,800 learners \citep{HPCCarpentry_Progress_2025}. The CaRCC \emph{People Network} hosts monthly role-based calls on research
computing facilitation, drawing 400–600 participants per session, while
NSF’s \emph{ACCESS} project is funding a coordinated national training
curriculum that bundles asynchronous MOOC content with hands-on bootcamps \citep{CaRCC_People_2024,ACCESS_Training_2025}.

\textbf{Barrier.}  Demand still outstrips supply: Hyperion Research
projects a shortfall of 2,500–3,000 HPC professionals in North America by
2030, driven by retirements and the AI-induced rise in system complexity \citep{Hyperion_Workforce_2024}. Retention is further hampered by salary competition with cloud vendors and
national laboratories, leading to 18-month median tenure for junior
system administrators at R1 universities.

\section{Conclusion}
  \subsection{Summary of Insights}
  This survey has undertaken a comprehensive review of HPC infrastructures across the U.S. academic landscape, benchmarking university-operated systems against Department of Energy leadership-class facilities and industrial hyperscale AI clusters. The findings underscore a persistent and widening gap between academic and non-academic sectors in both computational capacity and architectural advancement. While top-tier university systems such as Frontera and HiPerGator-AI have made commendable strides, the typical campus HPC cluster remains limited by funding cycles, heterogeneity of resources, and fragmented governance models. Most university clusters peak in the low-petascale range and rarely feature in the global TOP500’s upper echelon, in stark contrast to DOE systems like El Capitan (1.74 EF) and industrial AI supercomputers such as NVIDIA’s Eos.

Architecturally, the divergence is also notable. Universities tend to maintain CPU-heavy, general-purpose clusters with relatively modest GPU density, whereas national labs and industrial hyperscalers have rapidly adopted GPU-centric and custom-accelerated designs optimized for both simulation and AI workloads. Interconnect technologies also differ, with universities often relying on cost-effective Ethernet-based RoCE networks, while DOE and industrial systems deploy high-speed fabrics such as Slingshot-11 and Quantum-2 InfiniBand. Furthermore, energy efficiency has become a key differentiator; while some university systems like HiPerGator-AI achieve commendable GFLOPS/W ratings, they still lag behind industrial testbeds and DOE’s exascale systems in power efficiency and sustainability engineering.

Nevertheless, the academic sector shows promise in emerging paradigms such as decentralized reinforcement learning and heterogeneous GPU harvesting. Frameworks like INTELLECT-2 demonstrate that even fragmented campus hardware pools can support scalable, fault-tolerant AI training if properly federated and governed. Innovations in scheduling (e.g., MISO, MIGER) and buy-in governance models further offer pathways to extend and optimize limited campus resources.
  \subsection{Future Work }
To ensure continued relevance and policy impact, this survey should evolve into a recurring effort—ideally conducted on an annual basis. Regular tracking of university HPC upgrades, DOE investments, and industrial AI infrastructure expansions will enable more precise monitoring of sectoral trends, resource inequalities, and emerging best practices. Such annual snapshots can help university administrators and funding agencies make data-driven decisions, benchmark progress, and advocate for equitable federal investment in academic HPC.

In addition, expanding the scope to include international university HPC centers—such as those in Europe (e.g., PRACE, Jülich Supercomputing Centre) or Asia (e.g., RIKEN, Tsinghua)—would provide a valuable global context. Comparative analyses could reveal alternative governance models, architectural innovations, and funding mechanisms that may inform more effective strategies in the U.S. landscape. Ultimately, fostering global benchmarks and collaborative tracking will support a more inclusive and competitive international HPC research ecosystem.
\appendix

\bibliographystyle{unsrt}
\bibliography{mybib}
\end{document}